\documentclass[conference]{IEEEtran}
\IEEEoverridecommandlockouts
\usepackage{cite}
\usepackage{acronym}
\usepackage{verbatim}

\usepackage[pdftex]{graphicx}
\DeclareGraphicsExtensions{.pdf,.jpeg,.png}

\usepackage{xspace}
\usepackage[cmex10]{amsmath}
\usepackage{algorithm}
\usepackage[noend]{algcompatible}
\usepackage[tight,footnotesize]{subfigure}
\usepackage{url}
\usepackage{booktabs}
\usepackage{tikz}
\usetikzlibrary{automata}

\newcommand{\UPON}{\textbf{upon }}
\newcommand{\evnowifi}{\textsf{EV\_NO\_WIFI}\xspace}
\newcommand{\evshortwifi}{\textsf{EV\_SHORT\_WIFI}\xspace}
\newcommand{\evlongwifi}{\textsf{EV\_LONG\_WIFI}\xspace}

\acrodef{ABPS}{Always Best Packet Switching}
\acrodef{AP}{Access Point}
\acrodef{CN}{Correspondent Node}
\acrodef{CTMC}{Continuous Time Markov Chain}
\acrodef{MN}{Mobile Node}
\acrodef{MT}{Mobile Terminal}
\acrodef{NIC}{Network Interface Card}
\acrodef{SIP}{Session Initiation Protocol}
\acrodef{WiGLE}{Wireless Geographic Logging Engine}

\begin{document}

\title{Walking with the Oracle: Efficient Use of Mobile Networks through Location-Awareness\thanks{This work has been partially funded by the Italian Ministry of Education, University and Research (MIUR) through project STEM-Net (Prot. 2009S7RLXY\_003 - Year 2009). A revised version of this paper appears in the proceedings of Wireless Days 2012, Novembre 21--23 2012, Dublin, Ireland.}}

\author{\IEEEauthorblockN{Stefano Ferretti, Vittorio Ghini, Moreno Marzolla, Fabio Panzieri\\}
\IEEEauthorblockA{Department of Computer Science, University of Bologna\\
Mura A. Zamboni 7, I-40127 Bologna, Italy\\
\{sferrett, ghini, marzolla, panzieri\}@cs.unibo.it}
}

\maketitle

\begin{abstract}
Always Best Packet Switching (ABPS) is a novel approach for wireless
communications that enables mobile nodes, equipped with multiple
network interface cards (NICs), to dynamically determine the most
appropriate NIC to use. Using ABPS, a mobile node can seamlessly
switch to a different NIC in order to get better performance, without
causing communication interruptions at the application level. To make
this possible, NICs are kept always active and a software monitor
constantly probes the channels for available access points. While this
ensures maximum connection availability, considerable energy may be
wasted when no access points are available for a given NIC. In this
paper we address this issue by investigating the use of an ``oracle''
able to provide information on network availability. This allows to
dynamically switch on/off NICs based on reported availability, thus
reducing the power consumption. We present a Markov model which allows
us to estimate the impact of the oracle on the ABPS mechanism: results
show that significant reduction in energy consumption can be achieved
with minimal impact on connection availability. We conclude by
describing a prototype implementation of the oracle based on Web
services and geolocalization.
\end{abstract}

\section{Introduction}

The current generation of mobile devices (smartphones, tablets,
laptops) is frequently equipped with several~\acp{NIC} (e.g.~3G,
WiFi). The user must manually select which interface to use at any
given time, according to the availability of~\ac{AP} within
communication distance. In order to make better use of multiple
interfaces, we recently proposed the~\ac{ABPS}
mechanism~\cite{ngmast,greencom,GhiniJSS}. \ac{ABPS} keeps all
interfaces active at the same time, and uses the best one for data
transmission according to predefined criteria (e.g., availability or
connection quality). Furthermore, \ac{ABPS} can automatically switch
to a different~\ac{NIC} when the current connection terminates. The
interesting feature of~\ac{ABPS} is that all this is carried out
transparently to the applications and without any user intervention.
The idea is that a software component running in the~\ac{MN}
dynamically selects a ``preferred'' \ac{NIC} to use based on the
available access points. If the performance of that preferred~\ac{NIC}
degrades or the connection fails, the~\ac{MN} switches to a different
\ac{NIC}. Specifically, for each datagram, the \ac{MN} selects the
best \ac{NIC} to use.

A prototype implementation of~\ac{ABPS} is available~\cite{GhiniJSS},
and supports interactive multimedia services based on SIP/RTP/RTCP.
In~\cite{ngmast} we have shown that our approach provides better
performance in terms of availability, reliability and throughput, with
respect to conventional wireless communication
services. In~\cite{greencom} we studied the power consumption
of~\ac{ABPS}: our analysis revealed that, under some realistic
conditions, \ac{ABPS} requires less energy than traditional
single~\ac{NIC} systems, and other classic~\ac{SIP}-based
multi-\ac{NIC} communication approaches.  However, a particular aspect
of~\ac{ABPS} is that a software module at the~\ac{MN} continuously
monitors all the~\acp{NIC} and probes the communication channels
looking for access points to connect to. When there are no access
points for a specific~\ac{NIC} in the current area, keeping that
interface active has no practical benefit and negatively affects
the~\ac{MN} battery duration \cite{lee}.

In this paper, we propose an extension to~\ac{ABPS} based on an
additional software component--called ``oracle''--that is responsible
for predicting the availability (or lack of) of access points for the
available interfaces. Using this information, the~\ac{MN} can decide
to switch off a~\ac{NIC} for which no access points are currently
available. We describe a prototype implementation of the oracle based
on a software daemon which periodically queries the WiGLE Web
service~\cite{wigle} to locate WiFi networks within the
current~\ac{MN} location. 

To evaluate the impact of the oracle on connection availability, power
consumption and throughput, we define a performance model based
on~\acp{CTMC} with rewards. The model allows us to efficiently analyze
the system in different scenarios; such analysis would be impractical
using conventional measurement-based techniques. Results show
that~\ac{ABPS} enhanced with the oracle can significantly reduce the
power consumption of the~\ac{MN}, with just a marginal reduction of
availability and throughput with respect to the classic~\ac{ABPS}. We
believe that the performance model is useful by itself, as it can be
easily extended to describe multiple alternative implementations of
the oracle, allowing ``what-if'' analyses to be carried out
efficiently.

This paper is organized as follows. Section~\ref{sec:module} briefly
introduces the~\ac{ABPS} mechanism and describes the extension based
on the oracle. In Section~\ref{sec:prototype} we describe an initial
prototype implementation of the oracle based on the WiGLE service.
Section~\ref{sec:model} describes the Markov model exploited to
analyze the approach. Such model is evaluated in
Section~\ref{sec:eval}. Finally, concluding remarks are given in
Section~\ref{sec:conc}.

\section{The System}\label{sec:module}

\subsection{The Basic ABPS Mechanism}

The idea behind the basic~\ac{ABPS} mechanism is simple: during
transmission and reception, the~\ac{MN} can use all the
available~\acp{NIC} simultaneously, differentiating the choice of the
actual~\ac{NIC} on a datagram-by-datagram basis. This is carried out
transparently to both the application running in the~\ac{MN} and its
peer at the~\ac{CN}; thus, applications can continue to employ classic
end-to-end application-layer protocols such as TCP regardless
of~\ac{ABPS}.

In order to let the data flow through different~\acp{NIC}, and deliver
the related contents as a single flow to the application, additional
software components are needed.  Specifically, \ac{ABPS} requires the
\emph{ABPS Client Proxy} and the \emph{ABPS Server Proxy}
(see~\cite{GhiniJSS} for details).  The Client Proxy runs on
the~\ac{MN} and maintains the seamless multi-path communication
channel between the~\ac{MN} and the Server Proxy running on
the~\ac{CN}. The Server Proxy operates as a relay between the~\ac{MN}
and the~\ac{CN}, i.e.~it collects all the datagrams coming from
the~\ac{MN} via different~\acp{NIC}, and sends them to the application
running on the~\ac{CN} (application which is unaware of the presence
of the~\ac{ABPS} Server Proxy).

\subsection{An Oracle to Predict Network Availability}

The~\ac{ABPS} uses multiple~\acp{NIC} at the same time, and assumes
that all of them are always active (although possibly not always
connected to an~\ac{AP}). This ensures maximum connection
availability~\cite{ngmast}, and in some cases the use of \ac{ABPS} can
even reduce the power consumption with respect to a classic SIP based
approach~\cite{greencom}.  However, it remains true that the monitor
at the~\ac{ABPS} client proxy continuously probes the communication
channels looking for new access points to connect to. When there are
no access points available in the current area, this procedure is
useless and wastes energy, which is a particularly serious issue for
battery-powered mobile devices. To avoid such undesired situation, we
need some component (the ``oracle'') providing information on the
available access points for a given~\ac{NIC}, without explicitly
probing the communication channel. If the oracle reports that, for
some~\ac{NIC}, no access point will be available for a sufficiently
long period of time, than that~\ac{NIC} can be safely switched off.
On the other hand, if the oracle detects that the~\ac{MN} is entering
an area where access points are available, the~\ac{NIC} can be
reactivated.

\subsection{The Algorithm}

Before describing how the oracle may be implemented, let us focus in
detail on how the oracle can be used by considering a practical
scenario. We start by observing that most current smartphones are
typically equipped with both a WiFi and a 3G~\ac{NIC} (e.g.~UMTS). In
general, 2G/3G network coverage is quite ubiquitous, while WiFi access
points tend to be concentrated within densely populated areas (e.g.,
city centers). On the other hand, WiFi provides better bandwidth,
lower latency and lower energy consumption~\cite{greencom}, therefore
UMTS networks should always be used as a fallback when no usable WiFi
connection is available.

However, if the~\ac{MN} is entering an area where the WiFi connection
is likely to be available for a short time only (e.g., because
the~\ac{MN} is approaching an isolated access point), switching off
the UMTS~\ac{NIC} would not be a good idea since that connection will
be needed again in a short time. On the other hand, if the~\ac{MN} is
entering an area where WiFi coverage is guaranteed for a sufficiently
long time, then it is reasonable to switch UMTS off in order to save
energy.

We assume that the oracle has a list of available WiFi access points
in the current area; also, the oracle knows the current position of
the~\ac{MN}, and can also infer the future destination of the node
(see Section~\ref{sec:prototype} for a possible implementation).
Algorithm~\ref{alg:oracle} describes the high level behavior of the
oracle. Based on the list of available WiFi networks, the oracle
generates one of these events:

\begin{itemize}
\item \evnowifi: this event is generated with no WiFi access point is
  available; therefore, the WiFi~\ac{NIC} can be turned off so that
  connectivity is provided by the 2G/3G network only (lines
  \ref{alg:wifi_down_begin}--\ref{alg:wifi_down_end}).

\item \evshortwifi: this event is generated with a WiFi connection is
  going to be available for a short period of time; in this case both
  the UMTS and WiFi~\acp{NIC} are activated (lines
  \ref{alg:abps_begin}--\ref{alg:abps_end}).

\item \evlongwifi: this event is generated when a WiFi connection is
  going to be available for a longer period of time; in this case
  the~\ac{MN} can safely switch off UMTS (lines
  \ref{alg:umts_down_begin}--\ref{alg:umts_down_end}).
\end{itemize}

\begin{algorithm}[t]
\caption{ABPS Oracle} \label{alg:oracle}
\begin{small}
\begin{algorithmic}[1]
\STATE \UPON \evnowifi \hfill\COMMENT{no available WiFi network}\label{alg:wifi_down_begin}
\STATE \qquad $active_{WiFi} \gets false$
\STATE \qquad $active_{UMTS} \gets true$ \label{alg:wifi_down_end}
\STATE \UPON \evshortwifi \hfill\COMMENT{WiFi available for a short period}\label{alg:abps_begin}
\STATE \qquad $active_{WiFi} \gets true$
\STATE \qquad $active_{UMTS} \gets true$ \label{alg:abps_end}
\STATE \UPON \evlongwifi \hfill\COMMENT{WiFi available for a long period}\label{alg:umts_down_begin}
\STATE \qquad $active_{WiFi} \gets true$
\STATE \qquad $active_{UMTS} \gets false$ \label{alg:umts_down_end}
\end{algorithmic}
\end{small}
\end{algorithm}

The events (and associated new system configuration) generated by the
oracle is exploited by the~\ac{ABPS} Client Proxy as shown in
Algorithm~\ref{alg:client}.  When a given~\ac{NIC} is active, based on
the oracle prediction, and if an access point is detected for
that~\ac{NIC}, then the Client Proxy starts a setup procedure to
activate a connection and exploit it.  Upon successful connection
setup, the Client Proxy includes that interface in a list of~\acp{NIC}
available to transmit data. When a connection drops, the
associated~\ac{NIC} is removed from the list of active interfaces.

Based on this choice on the~\ac{NIC} to use, the communication between
the~\ac{MN} and its~\ac{CN} is as follows. The Client Proxy intercepts
and manages data coming from/to the application (lines
\ref{alg:comm_start}--\ref{alg:comm_end}). When there is data to
transmit, the Client sends those data to the Server Proxy via the
preferred~\ac{NIC} in use.  As multiple networks can be used during
the communication, it is possible that the transmitted datagrams are
received out of order. To this end, a sequence numbering scheme is
used that enables the receiver Server Proxy to order the received
datagrams and discard possible duplicates.  Finally, the Client Proxy
exploits ACKs to identify if some datagram is to be
re-transmitted. When a timeout for the reception of an ACK occurs, the
Client Proxy tries to retransmit that datagram through an
alternative~\ac{NIC} among those active at that time.

\begin{algorithm}[t]
\caption{ABPS Client Proxy} \label{alg:client}
\begin{small}
\begin{algorithmic}[1]
\STATE \hfill\COMMENT{NICs' management}\label{alg:man_start}
\STATE \UPON available connection for $NIC \wedge active_{NIC}$ 
\STATE \qquad \textsc{setup}($NIC$)
\STATE \UPON successful setup for $NIC \wedge active_{NIC}$ 
\STATE \qquad $availableNICs$.\textsc{add}($NIC$)
\STATE \qquad $inUseNIC \gets$ \textsc{selectPreferred}($availableNICs$)
\STATE \UPON disconnection for $NIC$
\STATE \qquad $availableNICs$.\textsc{remove}($NIC$)\label{alg:man_end}
\STATE \qquad $inUseNIC \gets$ \textsc{selectPreferred}($availableNICs$)
\STATE \hfill\COMMENT{Communication}\label{alg:comm_start}
\STATE \UPON new $data$ to transmit 
\STATE \qquad \textsc{send}($data, inUseNIC$)
\STATE \qquad \textsc{setTimeout}($data$)
\STATE \UPON $data$ received from ABPS server proxy
\STATE \qquad \textsc{pushOrderedBuffer}($data$,$bufferToApp$)
\STATE \qquad \textsc{sendOrderedData}($bufferToApp$)
\STATE \UPON ACK for $data$ received
\STATE \qquad \textsc{removeTimeout}($data$)
\STATE \UPON timeout ACK for $data$
\STATE \qquad $NIC \gets availableNICs$.\textsc{selectAlternativeNIC}()
\STATE \qquad \textsc{send}($data, NIC$)
\STATE \qquad \textsc{setTimeout}($data$)\label{alg:comm_end}
\end{algorithmic}
\end{small}
\end{algorithm}

The~\ac{ABPS} Server Proxy runs on a remote host, separate from that
of the Client Proxy. The Server Proxy operates basically as a relay
between the~\ac{MN} and its~\ac{CN}.  In practice, it is convenient to
place the ABPS Server Proxy on a fixed node located outside any
firewall and NAT system.  Then, the~\ac{ABPS} Client and Server
Proxies collaborate and adopt policies for load balancing and
recovery, in order to maximize throughput and minimize loss rate and
economic costs.  In particular, the~\ac{ABPS} Server Proxy collects
all the datagrams coming from the~\ac{MN}, via different~\acp{NIC} and
sends them to the~\ac{CN} (which is unaware of the presence of
the~\ac{ABPS} Server Proxy).

\section{Prototype Implementation}\label{sec:prototype}

\subsection{WiFi Networks Detection with WiGLE}

We have developed a prototype software module that is in charge of
periodically querying the WiGLE Web service to locate WiFi networks
within the area of the mobile user.  \acf{WiGLE} is a submission-based
Web catalog of wireless networks~\cite{wigle}.  Such service provides
location and information of wireless networks world-wide; through a
java application or a Web browser it is possible to map, query and
update the database. It allows to obtain statistics on the performance
of networks, and for instance, it is possible to ask only for free (or
commercial) networks.  We have developed a Python script in charge of
retrieving from~\ac{WiGLE} a list of WiFi access points available in
the area of the~\ac{MN}, based on its geographical position, together
with some related information.

To obtain the actual position of the \ac{MN}, a positioning system
must be utilized, i.e.~GPS. It has been shown that the power
consumption related to the use of a GPS is noticeably lower than that
due to the use of a \ac{NIC} \cite{Carroll:2010}, but in any case the
evaluation described in Section~\ref{sec:eval} explicitly takes into
account the additional energy consumption of the GPS module and the
associating processing cycles. It is worth noticing that many apps
running on users' smartphones require the use of GPS to locate points
of interests~\cite{FerrettiG09}. Hence, it would be reasonable to
assume that GPS is likely to be active anyway even without the
execution of the oracle. In this case, the geolocalizaion does not
introduce additional power consumption.

\subsection{Use Cases}

We report a couple of examples in order to show the benefits of having
an oracle that provides the list of WiFi networks, and also some main
related issues.

\begin{figure}[t]
   \centering
   \includegraphics[width=.9\columnwidth]{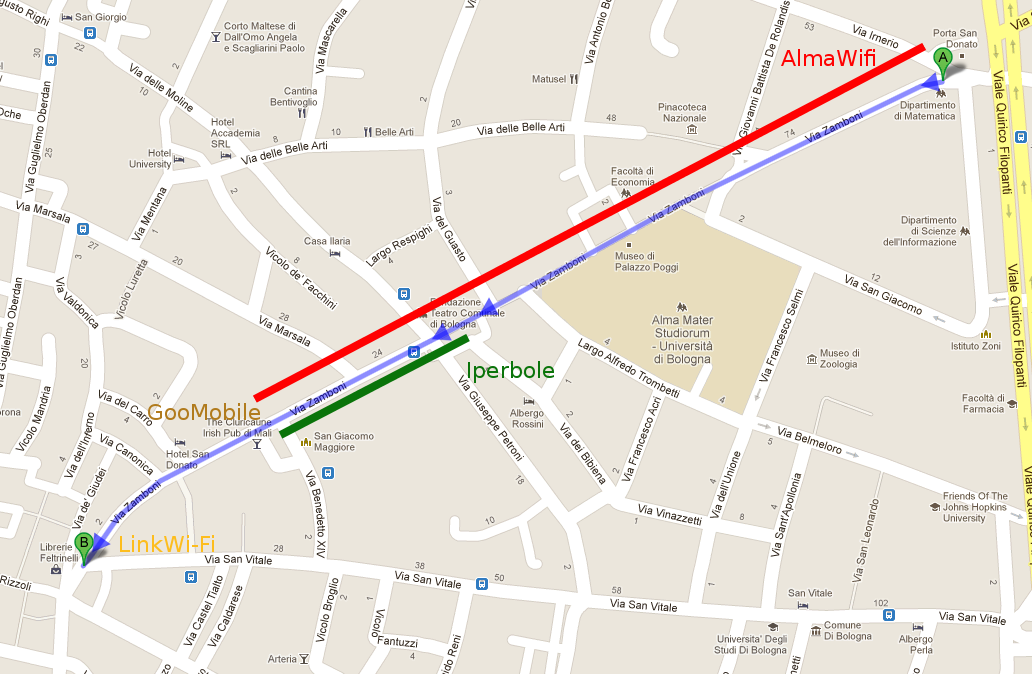}
   \caption{WiFi networks in Via Zamboni, Bologna (Italy)}
   \label{fig:zamboni}
\end{figure}

Let consider a mobile user that covers the path depicted in
Fig.~\ref{fig:zamboni}. The path is Via Zamboni in Bologna (Italy),
the main street in the city center where several departments of the
University of Bologna are located. In this area there are many active
WiFi access points. In particular, there are two main networks
available for students, i.e.~the AlmaWiFi network, which is the main
WiFi network for all students, faculty members and employees of the
University of Bologna, and the Iperbole civic network, available to
all citizens. For each point in the path, the oracle provides a list
of available access points.  For the sake of a clearer presentation,
we report only the essid of the access points with the best signal in
six main points of the path, i.e.~those corresponding to a square. The
red line in the Figure corresponds to the area in the path where there
is a network coverage provided by the AlmaWiFi Network; the green line
corresponds to the part of the path that is covered by some access
points of the Iperbole network. Other networks are available in the
network coverage of a single access point.

This example allows us to make some important considerations. Knowing
that a main part of the area is covered by a single WiFi network
(AlmaWiFi), the network manager of the~\ac{MN} might decide to switch
off its UMTS interface.  Obviously, during the path the mobile
terminal will perform horizontal handover through different access
points of the same network, but this is accomplished at the datalink
layer, transparently to the transport and application layers.  Then,
once the user becomes in proximity of the end of the AlmaWiFi network
coverage, the UMTS interface should be activated, so as to allow
the~\ac{MN} to exploit that interface during the handover from a WiFi
network to another one, or if no WiFi networks are available or open
to the user.

\begin{figure}[t]
   \centering
   \includegraphics[width=.9\columnwidth]{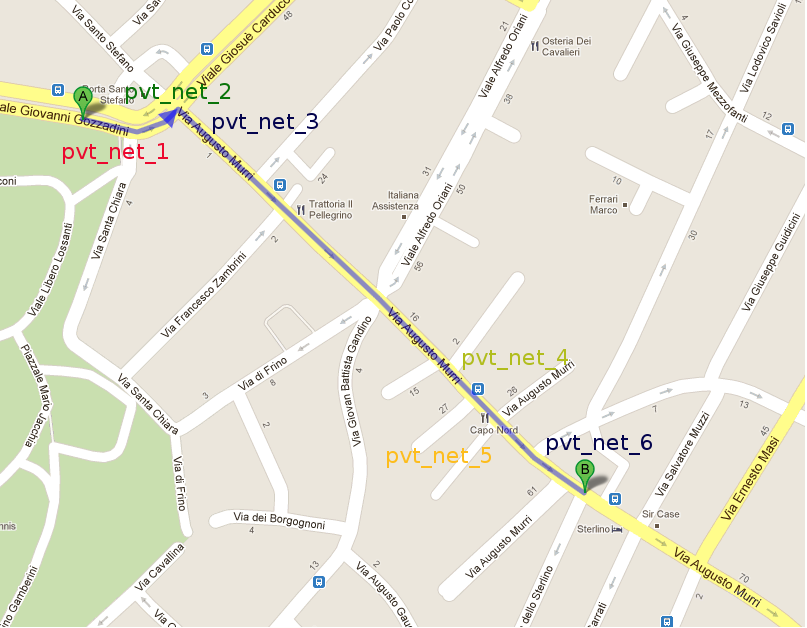}
   \caption{WiFi networks in Via Murri, Bologna (Italy)}
   \label{fig:margherita}
\end{figure}

Fig.~\ref{fig:margherita} shows a different path, starting from
Giardini Margherita, the main park in Bologna, and then straight to
via Murri, a street originating from the entrance of the park. In this
case, the oracle reports that there are private networks only (we do
not show them in the picture); moreover, the access points are located
at the beginning and the end of the path. Apparently, no WiFi networks
are available in the middle of the path. In this context, the best
strategy is to keep the UMTS~\ac{NIC} always active; the WiFi
interface can be switched on when the~\ac{MN} is in proximity of the
beginning and the end of the path, so that it can probe the available
networks and see if these networks are open and available.

\section{Modeling ABPS and the Oracle}\label{sec:model}

In this section we describe a performance model for~\ac{ABPS} with
oracle; the model is based on Markov chains with rewards, and extends
previous works on modeling the plain~\ac{ABPS}
mechanism~\cite{greencom,ngmast}. Our model can be solved very
quickly, and can be used to evaluate the proposed approach under
different parameter settings. We use the model to estimate the impact
of the oracle on the availability, power consumption and throughput
of~\ac{ABPS}. The insights provided by the model are being used to
drive current and future development of the oracle-enhanced \ac{ABPS}
prototype.\medskip

\paragraph*{Markov chains} 
We start by giving a very brief introduction to Markov chains
(see~\cite{bolch} for full details). A stochastic process $\{X(t), t
\geq 0\}$, defined over the discrete state space $\{1, \ldots, N\}$,
is a~\ac{CTMC} if the probability that the system is in state
$X(t_{n+1})$ at time $t_{n+1}$ only depends on the previous state
$X(t_n)$ at time $t_n$, for any $t_n < t_{n+1}$. A~\ac{CTMC} can
therefore be fully defined in term of an \emph{infinitesimal generator
  matrix} $\mathbf{Q} = [Q_{i,j}]$, where $Q_{i, j}$ is the transition
rate from state $i$ to state $j \neq i$. Given the infinitesimal
generator matrix and the initial state occupancy probability, we can
compute the state occupancy probability vector $\boldsymbol{\pi}(t) =
\left(\pi_1(t), \pi_2(t), \ldots, \pi_N(t) \right)$ at time $t$,
$\pi_i(t)$ being the probability that the system is in state $i$ at
time $t \geq 0$. Under certain conditions~\cite{bolch}, there exists a
stationary state occupancy probability $\boldsymbol{\pi} = \lim_{t
  \rightarrow +\infty} \boldsymbol{\pi}(t)$, which is independent from
the initial configuration of the Markov chain.  Of particular interest
for our analysis are \emph{Markov reward models}. We associate to each
state $i \in \{1, \ldots, N\}$ a nonnegative reward $r_i$. Rewards
have the following meaning: for each period of duration $dt$ spent in
state $i$, the total accumulated reward increases by $r_i dt$.

\begin{figure}[t]
  \centering%

  \subfigure[UMTS model\label{fig:model:umts}] {
    \begin{tikzpicture}[->,thick,>=stealth,node distance=3cm,scale=.7,transform shape]
      \tikzstyle{every state} = [fill=white];
      \node[state] (d) {$d_U$};
      \node[state] (s) [right of=d] {$s_U$};
      \node[state] (c) [right of=s] {$c_U$};
      \node[state] (f) [right of=c] {$f_U$};
      \begin{scope}[black!40]        
        \node[state] (off) [below of=d]  {$0_U$};
      \end{scope}
      
      \path
      (d) edge             node[above] {$\alpha_U$} (s)
      (s) edge             node[above] {$p_U \beta_U$} (c)
      (s) edge[bend left]  node[midway,below] {$(1-p_U) \beta_U$} (d)
      (c) edge             node[above] {$\gamma_U$} (f)
      (f) edge[bend right] node[midway,above] {$\mu_U$} (d);  

      \begin{scope}[black!40]
        \path
        (off) edge[bend left] node[left] {$[U1]$} (d) 
        (d) edge[bend left] node[right] {$[U0]$} (off)
        (s) edge[bend left] node[near start, below right] {$[U0]$} (off)
        (c) edge[bend left] node[near start, below right] {$[U0]$} (off)
        (f) edge[bend left] node[near start, below right] {$[U0]$} (off);
      \end{scope}

  \end{tikzpicture}}

  \subfigure[WiFi model\label{fig:model:wifi}] {
    \begin{tikzpicture}[->,thick,>=stealth,node distance=3cm,scale=.7,transform shape]
      \tikzstyle{every state} = [fill=white];
      \node[state] (d) {$d_W$};
      \node[state] (s) [right of=d] {$s_W$};
      \node[state] (c) [right of=s] {$c_W$};
      \node[state] (f) [right of=c] {$f_W$};
      \begin{scope}[black!40]
        \node[state] (off) [below of=d]  {$0_W$};
      \end{scope}
      
      \path
      (d) edge             node[above] {$\alpha_W$} (s)
      (s) edge             node[above] {$p_W \beta_W$} (c)
      (s) edge[bend left]  node[midway,below] {$(1-p_W) \beta_W$} (d)
      (c) edge             node[above] {$\gamma_W$} (f)
      (f) edge[bend right] node[midway,above] {$\mu_W$} (d);  

      \begin{scope}[black!40]
        \path
        (off) edge[bend left] node[left] {$[W1]$} (d) 
        (d) edge[bend left] node[right] {$[W0]$} (off)
        (s) edge[bend left] node[near start, below right] {$[W0]$} (off)
        (c) edge[bend left] node[near start, below right] {$[W0]$} (off)
        (f) edge[bend left] node[near start, below right] {$[W0]$} (off);
      \end{scope}

  \end{tikzpicture}}

  \subfigure[Oracle model\label{fig:model:oracle}] {
    \begin{tikzpicture}[->,thick,>=stealth,node distance=3cm,scale=.7,transform shape]
      \tikzstyle{every state} = [fill=white,minimum size=1.2cm];
      \node[state] (umts) {$O_U$};
      \node[state] (umtswifi) [right of=umts] {$O_{UW}$};
      \node[state] (wifi) [right of=umtswifi] {$O_W$};
      
      \path
      (umts)     edge[bend left]  node[midway,above] {$[W1]\ \lambda_{U, UW}$} (umtswifi)
      (umtswifi) edge[bend left]  node[midway,below] {$[W0]\ \lambda_{UW, U}$}  (umts)
      (umtswifi) edge[bend left]  node[midway,above] {$[U0]\ \lambda_{UW, W}$} (wifi)
      (wifi)     edge[bend left]  node[midway,below] {$[U1]\ \lambda_{W, UW}$} (umtswifi);
  \end{tikzpicture}}

  \caption{Markov models of ABPS with oracle}~\label{fig:model}
\end{figure}
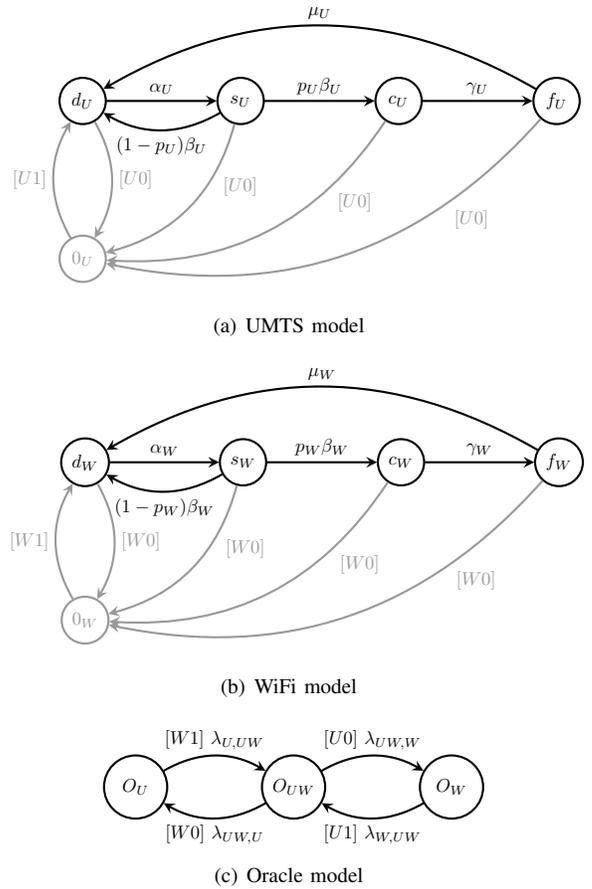

\paragraph*{Model Description}
The model is composed of three interacting~\acp{CTMC} as shown in
Fig.~\ref{fig:model}. The chain in Fig.~\ref{fig:model:umts}
represents the UMTS~\ac{NIC}, the one in Fig.~\ref{fig:model:wifi}
represents the WiFi~\ac{NIC}, and finally the chain in
Fig.~\ref{fig:model:oracle} represents the oracle.

The UMTS and WiFi models are identical (except for the parameter
values), so we focus on the UMTS chain shown in~\ref{fig:model:umts}.
The chain is defined over the state space $\{ 0_U, d_U, s_U, c_U, f_U
\}$; note that, for readability, we give symbolic names to states
instead of numeric IDs. State $0_U$ (\emph{off}) denotes that the
UMTS~\ac{NIC} is shut down. In state $d_U$ (\emph{disconnected}) the
UMTS interface is active and scans for networks to connect to. A
network is found after an average time $1 / \alpha_U$, and
the~\ac{NIC} enters state $s_U$ (\emph{setup}) where it tries to
connect to the newly found access point. A connection attempt has an
average duration of $1 / \beta_U$, after which it succeeds with
probability $p_U$ and fails with probability $(1-p_U)$. After a
successful connection, the UMTS~\ac{NIC} enters state $c_U$
(\emph{connected}); in this state data packets can be received and
transmitted. The average duration of a connection is $1 /
\gamma_U$. When the connection is lost, the UMTS interface enters
state $f_U$ (\emph{failed}). In this state no datagrams can be
received or delivered, but the~\ac{MN} is not yet aware of such
failure and continues to send data until it detects, eventually, that
the connection is not available and enters state $d_U$.

The oracle is modeled by the chain in
Fig.~\ref{fig:model:oracle}. State $O_U$ is entered when the \evnowifi
event is fired (see Algorithm~\ref{alg:oracle}); in this state only
the UMTS~\ac{NIC} is active. State $O_{UW}$ is entered when the
\evshortwifi event is fired; here both UMTS and WiFi are
active. Finally, state $O_W$ corresponds to event \evlongwifi where
only the WiFi~\ac{NIC} is active. In order to reduce the number of 
model parameters, we omitted direct transitions connecting
$O_W$ and $O_U$.

We can see that some of the transitions on Fig.~\ref{fig:model} are
labeled with \emph{actions} within square brackets. Actions are used
to force two chains to make transitions simultaneously, at a rate
equal to the product of the rates of synchronized transitions (when
the rate is omitted, it is assumed to be 1). We use synchronized
transitions to model the fact that the behavior of the oracle may
affect the behavior of the UMTS or WiFi chain. Actions $U1$ and $U0$
are used to switch on and off the UMTS interface. Action $U1$ is
executed when the transition $O_W \rightarrow O_{UW}$ is traversed,
forcing the transition $0_U \rightarrow d_U$ on the UMTS chain. On the
other hand, action $U0$ is executed when the oracle traverses the
transition $O_{UW} \rightarrow O_W$, forcing the UMTS chain to move to
state $0_U$ (\emph{off}) from whichever state it was in. Actions $W1$
and $W0$ have the same effect on the WiFi~\ac{NIC}.

The oracle has also another effect on the WiFi model, which is not
shown in the figure to reduce visual clutter. The value of $\gamma_W$,
which controls the average duration of WiFi connections, depends on
the state of the oracle chain. Recall that state $O_W$ is entered when
the event \evlongwifi is fired; therefore, in state $O_W$ the WiFi
connection should last longer. On the other hand, state $O_{UW}$ is
entered when the event \evshortwifi is fired, and in this case the
WiFi connection should have shorter duration. Let $T_W^+$ and $T_W^-$
be the mean duration of long and short WiFi connections. The
transition rate $\gamma_W$ from $c_W$ to $f_W$ is defined as follows:
if the oracle is in state $O_W$, then $\gamma_W = \gamma_W^+ = 1 /
T_W^+$, otherwise $\gamma_W = \gamma_W^- = 1 / T_W^-$. With these
rules the mean sojourn time in state $c_W$ is either $T_W^+$ or
$T_W^-$.

The complete model of~\ac{ABPS} with oracle is built by composing the
three individual chains shown in Fig.~\ref{fig:model}. The state space
is the set of triples $(S_\textrm{WiFi}, S_\textrm{UMTS},
S_\textrm{oracle})$, where $S_\textrm{UMTS} \in \{0_U, d_U, s_U, c_U,
f_U\}$, $S_\textrm{WiFi} \in \{0_W, d_W, s_W, c_W, f_W\}$ and
$S_\textrm{oracle} \in \{O_U, O_{UW}, O_W\}$. Not all configurations
are possible (e.g., when $S_\textrm{oracle} = O_W$ we know that UMTS
must be in state \emph{off}, hence $S_\textrm{UMTS} = 0_U$), therefore
the state space is actually smaller. The model has been implemented
using the PRISM model checker~\cite{prism} and can be found in the
Appendix.

\section{Analysis}\label{sec:eval}

We use the Markov model to study how the oracle affects three
quantities (\emph{availability}, \emph{power consumption} and
\emph{throughput}) with respect to the plain version of ABPS, where no
oracle is used. Plain ABPS is modeled by removing ``off'' states and
all transitions entering or exiting from them; basically, plain ABPS
is modeled by simply removing all gray elements from the models from
Fig.~\ref{fig:model}.

\paragraph*{Model Parameters}
We consider the following parameter values, which have been
empirically determined:
\begin{align*}
\alpha_U &= 1/6.024 & \alpha_W &= 1/7.5 \\
\beta_U &= 1/1.5 & \beta_W &= 1/1.5 \\
\gamma_U &= 1/500 & \gamma_W^- &= [1/5, 1/40] \\
&& \gamma_W^+ &= [1/40, 1/120] \\
\mu_U &= 1 & \mu_W &= 1 \\
p_U &= 0.99 & p_W &= 0.9 \\
\lambda_{U, UW} &= 1/30 & \lambda_{UW, W} &= \gamma_W^- / 2 \\
\lambda_{UW, W} &= \gamma_W^- / 2 & \lambda_{W, UW} &= \gamma_W^+
\end{align*}
Transition rates are defined as the inverse of experimentally measured
values. We set the duration of short WiFi connections (those for which
the oracle triggers the \evshortwifi event) in the range $[5, 40]$
seconds; long WiFi connections are set in the range $[40, 120]$
seconds. We keep the values of other parameters fixed, since the time
needed to set up a connection, or detect that a connection has failed,
are likely to be unaffected by the duration of WiFi connections.
Transition rates for the oracle have been set so that the expected
residence times in states $O_W$ and $O_{UW}$ matches the long and
short WiFi connection duration, respectively. Recall that the mean
residence time in a given state can be computed as the inverse of the
sum of rations of all outgoing transitions~\cite{bolch}. Therefore,
the mean residence time in state $O_W$ is $1 / \lambda_{W, UW} = 1 /
\gamma_W^+ = T_W^+$, and the mean residence time in state $O_{UW}$ is
$1 / (\lambda_{UW, W} + \lambda_{UW, U} ) = 1 / \gamma_W^- =
T_W^-$. The mean residence time in state $O_U$, where no WiFi
connection is available, has been empirically determined as $1 /
\lambda_{U, UW} = 30$ seconds.\medskip

\begin{figure*}[ht]
\centering%
\subfigure[Availability\label{fig:results:availability}]{\includegraphics[width=.32\textwidth]{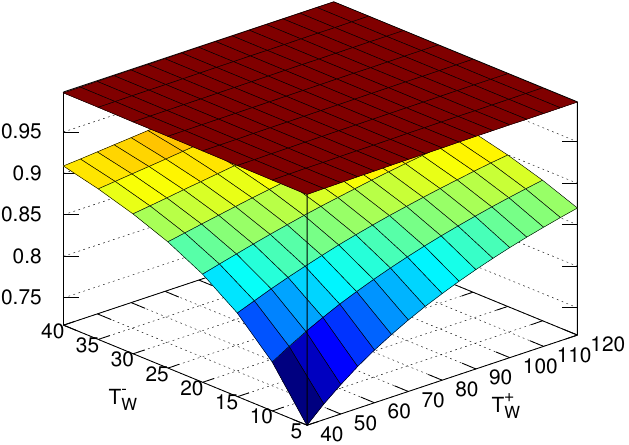}}\quad%
\subfigure[Power Consumption\label{fig:results:power}]{\includegraphics[width=.32\textwidth]{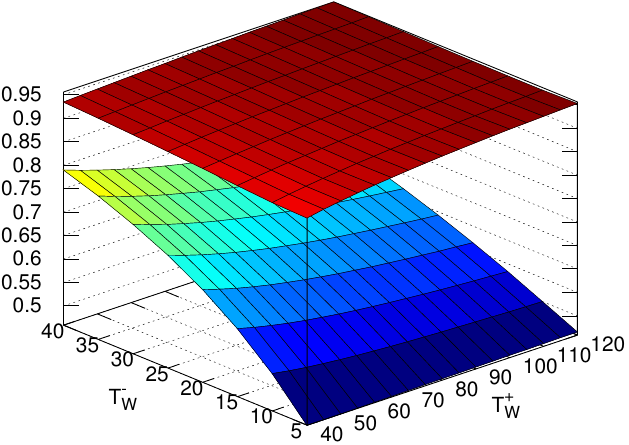}}\quad%
\subfigure[Throughput\label{fig:results:throughput}]{\includegraphics[width=.32\textwidth]{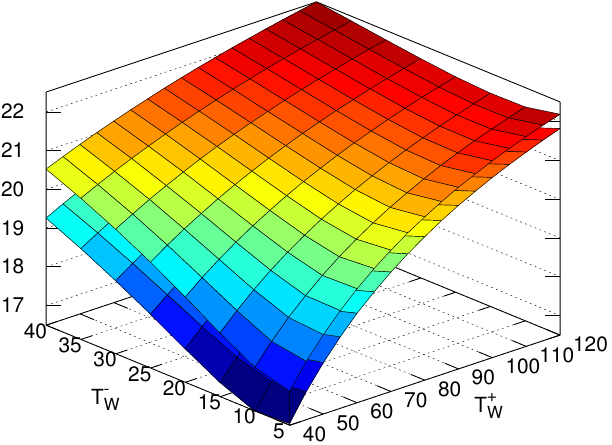}}
\caption{Model evaluation results for plain ABPS (top) and ABPS+oracle (bottom)}\label{fig:results}
\end{figure*}

\paragraph*{Connection Availability}
The connection availability is the long term fraction of actually
available service. Formally, the availability is the steady-state
probability that either the UMTS or WiFi~\acp{NIC} are connected
(i.e., in state $c_U$ or $c_W$,
respectively). Fig.~\ref{fig:results:availability} shows the
availability computed by PRISM as a function of $T_W^-$ and $T_W^+$.
(higher is better), for both plain~\ac{ABPS} and~\ac{ABPS} enhanced
with the oracle. Plain ABPS provides higher availability, especially
when the duration of WiFi connection is short. This can be explained
by observing that WiFi is more unreliable than UMTS (the average
duration of a WiFi connection is much lower than UMTS), hence the
oracle incurs the UMTS connection setup overhead each time the system
moves from state $O_W$ to $O_{UW}$. Plain ABPS does not incur this
overhead, since both interfaces are always active at the same time,
and the setup on one interface may be overlapped with the other one
being already connected. The availability improves if the value of
$T_W^-$ (the duration of short WiFi connections) is increased. The
reason is that longer values of $T_W^-$ give the interface being
activated in state $O_{UW}$ enough time to connect before the system
shuts down the other one. In practice, this result suggests that the
oracle should not trigger a \evshortwifi when the WiFi connection is
going to be available for a very short time only.\medskip

\paragraph*{Power Consumption}
In order to estimate the power consumption, we define a reward
structure in which each state is assigned a reward equal to the power
consumption (in Watts) of the~\acp{NIC} in that state.  When
both~\acp{NIC} are connected, we assume that only the fastest one
(WiFi) is used for data transmission. Therefore, in this case we
assume that the idle interface consumes $20\%$ of its peak
power. Additionally, for~\ac{ABPS}+oracle we increased the reward of
all states by $0.1W$, to account for the need to keep the GPS
connection active, and for executing the oracle application in
background. The contribution of other components of the~\ac{MN} is
assumed to be more or less the same in both scenarios (plain~\ac{ABPS}
and~\ac{ABPS}+oracle), so it is omitted in this analysis.  Rewards are
shown in Table~\ref{tab:energy}, and are based on data
from~\cite{energy_consumption}.

\begin{table}[t]
\centering%
\caption{Rewards for energy consumption estimation}\label{tab:energy}
\begin{tabular}{rccccc}
\toprule
 & \multicolumn{5}{c}{\bf State}\\
\cmidrule{2-6}
 & $0$ & $d$ & $s$ & $c$ & $f$ \\
\midrule
UMTS & 0.0 & 0.12 & 0.31 & 0.62 & 0.25\\
WiFi & 0.0 & 0.08 & 0.19 & 0.38 & 0.15\\
\bottomrule
\end{tabular}
\end{table}

Results are shown in Fig.~\ref{fig:results:power}: we observe that the
oracle significantly reduces the energy requirement of the~\ac{MN},
despite the fact that it uses the GPS device to get the current
position. We also observe that availability and power consumption are
correlated, as expected: higher availability requires more energy

\paragraph*{Throughput}
We finally evaluate the throughput provided by the two communication
mechanisms. Again, we use a reward structure in which we associate to
each state the average throughput provided by the~\ac{NIC} in that
state. This means that for the UMTS chain we associate a throughput of
$0.2$ Mbps to state $c_U$, and for the WiFi chain we associate a
throughput of $26$ Mbps to state $c_W$. When both interfaces are
connected at the same time, we only use WiFi for actual data
transmission. As can be seen in Fig.~\ref{fig:results:throughput}, the
oracle guarantees a throughput that is only marginally lower than that
provided by plain~\ac{ABPS}. Unsurprisingly, the duration of WiFi
connections plays a major role on the system throughput. When WiFi
connections last longer, the~\ac{MN} can make better use of that to
achieve higher transmission rates. This happens for both
plain~\ac{ABPS} and~\ac{ABPS} enhanced with the oracle.

\section{Conclusion}\label{sec:conc}

In this paper we proposed an extension of the~\ac{ABPS} communication
mechanism based on a software component (the oracle) which can predict
the availability of access points as the~\ac{MN} moves. This
information can be exploited to switch off the interface for which no
access points are going to be available in the near future, in order
to save energy. We described an actual prototype implementation of the
oracle based on the WiGLE service; in order to better evaluate the
impact of the oracle on~\ac{ABPS}, we defined a performance model
based on continuous-time Markov chains that is quite simple and can be
efficiently solved. We used the model to compare the connection
availability, steady-state power consumption and throughput of
plain~\ac{ABPS} and~\ac{ABPS} with oracle, for different scenarios.
The results show that the oracle can provide a significant reduction
of the power consumption with marginal reduction of availability and
throughput. The Markov model is currently being used to drive the full
implementation of the oracle-based~\ac{ABPS}, by providing quick
performance estimates of various implementation alternatives.

\appendix[PRISM models]

\subsection{Plain ABPS Model}

\begin{scriptsize}
\begin{verbatim}
// abps-plain.sm

ctmc

// Externally defined parameters
const double T_W_plus;
const double T_W_minus;

// UMTS model transition rates
const double alpha_U = 1/6.024;
const double beta_U = 1/1.5;
const double gamma_U = 1/600;
const double mu_U = 1.0;
const double p_U = 0.99;

// WiFi model transition rates
const double alpha_W = 1/7.5;
const double beta_W = 1/1.5;
const double gamma_W_plus = 1/T_W_plus;
const double gamma_W_minus = 1/T_W_minus;
const double mu_W = 1.0;
const double p_W = 0.9;

// Oracle transition rates
const double lambda_12 = 30; // gamma_U;
const double lambda_21 = 0.5*gamma_W_minus;
const double lambda_23 = lambda_21;
const double lambda_32 = gamma_W_plus;

// Power consumption for UMTS
const double e_U_0 = 0.0;
const double e_U_1 = 0.12;
const double e_U_2 = 0.31;
const double e_U_3 = 0.62;
const double e_U_4 = 0.25;

// Power consumption for WiFi
const double e_W_0 = 0.0;
const double e_W_1 = 0.08;
const double e_W_2 = 0.19;
const double e_W_3 = 0.38;
const double e_W_4 = 0.15;

// NIC throughput (Mbps)
const double Tput_U = 0.2; // UMTS
const double Tput_W = 26;  // WiFi


// Baseline power consumption
const double e_base = 0.0;

//
// UMTS interface definition
//
module umts
   // state enumeration:
   // 1 = disconnected
   // 2 = setup
   // 3 = connected
   // 4 = failed
   s_U : [1..4] init 1;

   [] s_U=1 -> alpha_U:(s_U'=2);
   [] s_U=2 -> beta_U*p_U:(s_U'=3) +
               beta_U*(1.0-p_U):(s_U'=1);
   [] s_U=3 -> gamma_U:(s_U'=4);
   [] s_U=4 -> mu_U:(s_U'=1);
endmodule

//
// WiFi interface definition
//
module wifi
   s_W : [1..4] init 1;

   [] s_W=1 -> alpha_W:(s_W'=2);
   [] s_W=2 -> beta_W*p_U:(s_W'=3) +
               beta_W*(1.0-p_W):(s_W'=1);
   [] s_W=3 -> (s_oracle = 3 ? 
               gamma_W_plus : 
               gamma_W_minus):(s_W'=4);
   [] s_W=4 -> mu_W:(s_W'=1);
endmodule

//
// Oracle model definition
//
module oracle
       // 1 = UMTS only
       // 2 = UMTS + WiFi
       // 3 = WiFi only
       s_oracle : [1..3] init 2;
       
       [] s_oracle=1 -> lambda_12:(s_oracle'=2);
       [] s_oracle=2 -> lambda_21:(s_oracle'=1);
       [] s_oracle=2 -> lambda_23:(s_oracle'=3);
       [] s_oracle=3 -> lambda_32:(s_oracle'=2);
endmodule

//
// reward structures
//

formula U_connected	= s_U=3;  // is UMTS connected?
formula U_not_connected = s_U!=3; // is UMTS NOT connected?
formula W_connected 	= s_W=3;  // is WiFi connected?
formula W_not_connected = s_W!=3; // is WiFi NOT connected?

// Power Consumption
rewards "energy"
    true: e_base; // add baseline power consumption
                  // to all states
    s_U=0: e_U_0;
    s_U=1: e_U_1;
    s_U=2: e_U_2;
    s_U=3: e_U_3;
    s_U=4: e_U_4;

    s_W=0: e_W_0;
    s_W=1: e_W_1;
    s_W=2: e_W_2;
    s_W=3: e_W_3;
    s_W=4: e_W_4;
endrewards

// Throughput 
rewards "throughput"
    W_connected & U_not_connected: Tput_W;
    U_connected & W_not_connected: Tput_U;	
    U_connected & W_connected    : Tput_W;
endrewards
\end{verbatim}
\end{scriptsize}

\subsection{Plain ABPS Property Specification}

\begin{scriptsize}
\begin{verbatim}
// abps-plain.csl

// Stationary energy consumption rate
"power": R{"energy"}=? [S]

// Stationary connection availability
"availability": S=? [ s_U = 3 | s_W = 3 ]

// Stationary throughput
"throughput": R{"throughput"}=? [S]
\end{verbatim}
\end{scriptsize}

\subsection{ABPS+Oracle Model}

\begin{scriptsize}
\begin{verbatim}
// abps-oracle.sm

ctmc

// Externally defined parameters
const double T_W_plus;
const double T_W_minus;

// UMTS model transition rates
const double alpha_U = 1/6.024;
const double beta_U = 1/1.5;
const double gamma_U = 1/600;
const double mu_U = 1.0;
const double p_U = 0.99;

// WiFi model transition rates
const double alpha_W = 1/7.5;
const double beta_W = 1/1.5;
const double gamma_W_plus = 1/T_W_plus;
const double gamma_W_minus = 1/T_W_minus;
const double mu_W = 1.0;
const double p_W = 0.9;

// Oracle transition rates
const double lambda_12 = 30;
const double lambda_21 = 0.5*gamma_W_minus;
const double lambda_23 = lambda_21;
const double lambda_32 = gamma_W_plus;

// Power consumption for UMTS
const double e_U_0 = 0.0;
const double e_U_1 = 0.12;
const double e_U_2 = 0.31;
const double e_U_3 = 0.62;
const double e_U_4 = 0.25;

// Power consumption for WiFi
const double e_W_0 = 0.0;
const double e_W_1 = 0.08;
const double e_W_2 = 0.19;
const double e_W_3 = 0.38;
const double e_W_4 = 0.15;

// NIC throughput (Mbps)
const double Tput_U = 0.2; // UMTS
const double Tput_W = 26;  // WiFi

//
// UMTS model definition
//
module umts
   // state enumeration:
   // 0 = off 
   // 1 = disconnected
   // 2 = setup
   // 3 = connected
   // 4 = failed
   s_U : [0..4] init 1;

   [] s_U=1 -> alpha_U:(s_U'=2);
   [] s_U=2 -> beta_U*p_U:(s_U'=3) +
               beta_U*(1.0-p_U):(s_U'=1);
   [] s_U=3 -> gamma_U:(s_U'=4);
   [] s_U=4 -> mu_U:(s_U'=1);
   
   [umts_0] true -> (s_U'=0);

   [umts_1] s_U=0 -> (s_U'=1);
endmodule

//
// WiFi model definition
//
module wifi
   s_W : [0..4] init 1;

   [] s_W=1 -> alpha_W:(s_W'=2);
   [] s_W=2 -> beta_W*p_U:(s_W'=3) +
               beta_W*(1.0-p_W):(s_W'=1);
   // If the oracle is in state 3, the mean
   // duration of WiFi connected state is
   // set to T_W_plus
   [] s_W=3 -> ( s_oracle = 3  ? 
                 gamma_W_plus  : 
                 gamma_W_minus ):(s_W'=4);
   [] s_W=4 -> mu_W:(s_W'=1);
   
   [wifi_0] true -> (s_W'=0);

   [wifi_1] s_W=0 -> (s_W'=1);
endmodule

//
// Oracle model definition
//
module oracle
   // State space enumeration:
   // 1 = UMTS only
   // 2 = UMTS + WiFi
   // 3 = WiFi only
   s_oracle : [1..3] init 2;
       
   [wifi_1] s_oracle=1 -> lambda_12:(s_oracle'=2);
   [wifi_0] s_oracle=2 -> lambda_21:(s_oracle'=1);
   [umts_0] s_oracle=2 -> lambda_23:(s_oracle'=3);
   [umts_1] s_oracle=3 -> lambda_32:(s_oracle'=2);
endmodule

//
// Definition of reward structures
//

formula U_connected	= s_U=3;  // is UMTS connected?
formula U_not_connected = s_U!=3; // is UMTS NOT connected?
formula W_connected 	= s_W=3;  // is WiFi connected?
formula W_not_connected = s_W!=3; // is WiFi NOT connected?

// Power Consumption
rewards "energy"
    true: 0.1;  // add baseline power consumption
                // to all states to account for
                // GPS usage and CPU cycles
    s_U=0: e_U_0;
    s_U=1: e_U_1;
    s_U=2: e_U_2;
    s_U=3: e_U_3;
    s_U=4: e_U_4;

    s_W=0: e_W_0;
    s_W=1: e_W_1;
    s_W=2: e_W_2;
    s_W=3: e_W_3;
    s_W=4: e_W_4;
endrewards

// Throughput 
rewards "throughput"
    W_connected & U_not_connected: Tput_W;
    U_connected & W_not_connected: Tput_U;	
    U_connected & W_connected    : Tput_W;
endrewards
\end{verbatim}
\end{scriptsize}

\subsection{ABPS+Oracle Property Specification}

\begin{scriptsize}
\begin{verbatim}
// abps-oracle.csl

// Stationary energy consumption rate
"power": R{"energy"}=? [S]

// Stationary connection availability
"availability": S=? [ s_U = 3 | s_W = 3 ]

// Stationary throughput
"throughput": R{"throughput"}=? [S]
\end{verbatim}
\end{scriptsize}



\end{document}